# Less (Data) Is More: Why Small Data Holds the Key to the Future of Artificial Intelligence

Ciro Greco[*], Andrea Polonioli[*] and Jacopo Tagliabue[*]
*Tooso Labs,*
*San Francisco, CA, U.S.A.*

Keywords: Artificial Intelligence, Big Data, Probabilistic Programming, Concept Learning, Machine Learning.

Abstract: The claims that big data holds the key to enterprise successes and that Artificial Intelligence (AI) is going to replace humanity have become increasingly more popular over the past few years, both in academia and in the industry. However, while these claims may indeed capture some truth, they have also been massively oversold, or so we contend here. The goal of this paper is two-fold. First, we provide a qualified defence of the value of less data within the context of AI. This is done by carefully reviewing two distinct problems for big data driven AI, namely a) the limited track record of Deep Learning (DL) in key areas such as Natural Language Processing (NLP), b) the regulatory and business significance of being able to learn from few data points. Second, we briefly sketch what we refer to as a case of "A.I. with humans and for humans", namely an AI paradigm whereby the systems we build are privacy-oriented and focused on human-machine collaboration, not competition. Combining our claims above, we conclude that when seen through the lens of cognitively inspired A.I., the bright future of the discipline is about less data, not more, and more humans, not fewer.

## 1 INTRODUCTION

The unreasonable effectiveness of data is possibly the greatest surprise coming out of the last twenty years of Artificial Intelligence (AI): pretty simple algorithms and tons of data seem to almost invariably beat complex solutions with small-to-none training set. In the seminal words of (Halevy, Norvig, and Pereira, 2009): "now go out and gather some data, and see what it can do".

The perfect storm has been set in motion by the convergence of the big data hype (Hagstroem et al 2017), the general availability of specialized hardware and scalable infrastructure, and some "computational tricks" (e.g. Hochreiter S., Schmidhuber S., 1997, Hinton et al, 2013): all together, they unlocked the Deep Learning (DL) Revolution and created a tremendous amount of business value (Chui et al 2018).

The A.I. wave is so disruptive that a great deal of commentators, practitioners (Radford et al 2019) and entrepreneurs (Musk 2017) inevitably started to wonder what is the place of *humans* in this new world: is A.I. going to replace humanity (in the world of Silicon Valley, Joy in 2001 was already stating that "the future doesn't need us")? In *this* position paper, we shall argue for two surprising perspectives: 1) the future of A.I. is about *less* data, not more; 2) human-machine collaboration is, at least for the foreseeable future, the only way to outpace humans and outsmart machines effectively.

The paper is organized as follows: *Section 2* contains a review of the current state of the A.I. landscape, with particular attention to the origins of the DL Revolution; the section casts some doubts on the general applicability of DL to language problems, drawing from theoretical considerations from academia and industry use cases in the space of Tooso. *Section 3* details a real use-case from the industry that is challenging for the DL paradigm, and outlines a different framework to tackle the problem; finally, *Section 4* concludes with remarks and roadmap for a new type of A.I., what we call "A.I. with humans and for humans."

---

[*] Authors have been listed alphabetically

## 2 THE RISE (AND FALL) OF "BIG DATA-DRIVEN" ARTIFICIAL INTELLIGENCE

The DL Revolution is conventionally linked to the seminal paper on ImageNet by (Krizhevsky, I., et al. 2012); as the 2019 Turing Award Ceremony makes clear, the theoretical impact of DL cannot be overstated (ACM 2019).

On the practical side, the recognition of DL potential has resulted into A.I. startups securing increasingly larger amounts of funding: between 2013 and 2017, Venture Capital (VC) investments in A.I. startups increased with a compound annual growth rate (CAGR) of about 36% (Su 2018). While A.I. and DL are by no means synonyms (DL being a subset of Machine Learning, which is itself a subset of A.I.), it's undeniable that DL is what mostly account for today's A.I. renaissance.

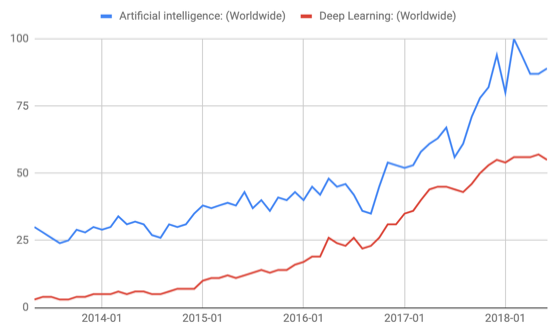

Figure 1: "A.I." and "Deep Learning" search trends, 2014-2018. Data source: Google Trends (https://www.google.com/trends).

To properly understand DL - and, more generally, the last twenty years of ML (Machine Learning) - it's crucial to grasp the relation between data and performance at the heart of all types of statistical learning.

Take a simple ML system for *spam filtering*. A message such as ''buy online cheap Viagra'' immediately stands out as spam-like to humans; on the other hand, ML systems generally solve problems like this one, by reframing them as statistical inference: how many times the text strings "buy online", "cheap" and "Viagra" are to be found together in a spam vs legit message? The more emails the algorithm has seen, the better it will be at making guesses. What takes a human few examples to learn, it takes ML systems millions: the bigger the training dataset, the easier is for the system to find patterns that are likely to occur when the system is asked to do new predictions - and this is why spam filters work so well: data is abundant and the task is a well- defined classification task.

Now that we understand the general ML paradigm of "intelligence as curve fitting" (Hartnett 2018), we can turn to the successes of DL, which literally redefined the meaning of "unreasonable effectiveness of data".

### 2.1 Technical and Hardware Advances

The idea that the brain, neuron-based, architecture could be a powerful inspiration for successful computational tools is pretty old. Even without considering the first days of neural systems (Rosenblatt 1957; Minsky, Papert 1969), the backpropagation algorithm at the heart DL is now more than 30 years old (Rumelhart *et al* 1986). Why are we experiencing the DL Revolution just now?

While neural networks and human brains are still very different things (Marblestone *et al* 2016), it was quickly realized that neural networks, as the brain, need many neurons to be effective. There is a catch though: the deeper the network, the more parameters need to be tuned; the more parameters, the more data points; the more data points, the more computational power.

At the cost of some simplification, we can identify two trends that converged to boost neural network performances in recent years: the big data trend made available datasets of unprecedented size; in turn, DL early successes pushed the market for widespread availability of dedicated hardware and software resources.

#### 2.1.1 Big Data (Data Collection, Data Storage, Data Processing)

The promises of statistically learning hidden patterns better than traditional ML can be only realized thanks to DL ability of leveraging massive amounts of data to tune their parameters (Hestness et al 2017). In recent years, data volumes have been constantly growing: if three exabytes of data existed in the mid Eighties, it was 300 by 2011; in 2016, the United States alone had more than 2 zettabytes (2000 exabytes) of data (Henke et al 2016). Increasing data availability put competitive pressure to get even more data and be able to store, retrieve, analyze massive datasets. The release of open source tools - such as Hadoop (Ghemawat et al 2003, Shvachko et al. 2010) and Spark (Zaharia et al 2010) - and widespread availability of cheap storage (especially though cloud growth) democratized the access to the world of big data for organizations of every kind and size (Hashem 2015).

### 2.1.2 Big Data (Data Collection, Data Storage, Data Processing)

DL algorithms are notoriously data hungry (Marcus 2018), but a lot of what happens inside a DL algorithm can be massively parallelized in commodity hardware, such as GPUs. Getting started with DL has indeed never been so easy: the availability of cloud-based (IDG 2018), pay-as-you-go GPUs (or even physical cards at a reasonable price range), coupled with the open source release of DL frameworks (e.g. Abadi et al 2015), resulted in the possibility of replicating state-of-the- art models from a common laptop (Tensorflow 2018). Interestingly enough, all the big tech players have been eager to release in the public domain the code for DL frameworks, knowing well that their competitive advantage is not so much in tooling, but in the vast amount of proprietary user data they can harvest.

## 2.2 Is A.I. Truly Riding a One-trick Pony?

Deep learning has led to important results in speech and image recognition and plays a key role in many current AI applications. However, the idea that DL represents the ultimate approach faces challenges and criticisms as well. While a thorough examination is out of the scope of the present article (see for example Marcus 2018, Lake et al 2016), we are content to list two reasons why we should remain open to new ideas:
1) in spite of deep learning's remarkable success in some domains of application, its track record in a key domain like natural language processing is far less outstanding;
2) data are clearly a key asset for enterprises, but big data are available to only a small subset of companies; truth to be told, even *within* enterprises, many use cases have severe constraints on data quantity and quality. Further, due to regulatory issues and compliance, even those who used to have access to them might face increasing difficulties.
Without presumption of completeness, the next subsections briefly elaborate on these points: taken all together, a critical appraisal of DL strongly points to the fact that "intelligence as curve fitting" cannot be the only paradigm for the next generation of A.I. systems.

### 2.2.1 Beginning of an Era… or End of One?

As incredible as the DL successes have been, there is also a general consensus that improvements are plautening fast (Chollet, 2017). Moreover, not all fields and benchmarks have been "disrupted" in similar ways, as it is easy to realize comparing the error rate progression, year after year, of the best deep learning model on visual task (ImageNet competition) vs language task (Winograd challenge).

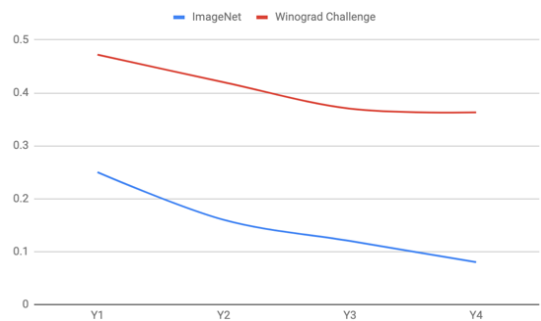

Figure 2: Error rate for the best performing deep learning models of the year in two standard challenges: ImageNet (vision-based challenge, 2011-2014) vs Winograd (language-based challenge, 2014-2017).

While both trends have been showing decreasing marginal gains since their inception, the error rate for the visual-based challenge (Deng et al 2009) reached human performances (~0.05) in four years; in the same timeframe, the error rate for the language-based challenge (Levesque 2011) did not even come close (~0.05). The problem with language is clearly more general than understanding and inference: state-of-the-art models in machine translation (Vaswani et al. 2017) achieve a sentence-level score of 28.4 (out of 100.0) for English-German (human performance is >75.0).

As nicely summarized by (Landgrebe and Smith 2019), deep neural nets typically outperform humans in numerical pattern recognition or context with a somewhat explicit knowledge that can be framed as games (Silver et al 2016). However, these types of situations are "highly restrictive, and none occurs where we are dealing with natural language input."

### 2.2.2 Life from the Trenches: Regulatory Changes and Data Constraints

A huge topic of discussion has been the claim that big data carry some possible harms for individuals whose data are being analyzed. Based on this claim, compliance and regulatory issues have recently become pressing concerns for enterprises dealing with huge amounts of data, especially after GDPR entered into force (Zarsky 2017). While the extent to which changes in the regulatory landscape will alter current big data practices for enterprises is still unclear, it is safe to bet that they will indeed lead to a shift, meaning that accumulating data will become riskier and more challenging.

Moreover, even in enterprises where generally data is abundant, there will be plenty of use cases that won't really fit the classic big data definition: sometimes data may be abundant in theory, but too dirty or simply lacking any meaningful label to be actionable; other times, data quantity will be indeed constrained by i) the use case at hand (e.g. doing hyper-personalization (Costa 2014) with few user data points) or ii) data distribution (e.g. given the power-law in query distribution for ecommerce search engines, more than 50% of queries involve dealing with very low frequency linguistic data (Brynjolfsson, et al 2011)). In general, even very successful ML algorithms struggle in making reliable inference with few data points (Lake et al 2015), while humans are exceptionally good at this (Markman 1989; Xu and Tenenbaum, 2007).

## 3 LEARNING LANGUAGE THROUGH HUMAN-MACHINE INTERACTION

One of the most celebrated aspects of DL - namely, that neural nets learn "automatically" which parts of the input are crucial for the outcome - turns out to be one of its greatest shortcomings: learning new things from scratch in a giant space of possible parameters makes it impossible to learn from few data points. A recent wave of cognitively inspired models (Goodman and Tenenbaum 2016) is starting to challenge the A.I. status quo, bringing a very different set of assumptions to the table: in the words of (Xu and Tenenbaum, 2007), "a structured hypothesis space can be thought of as (...) perhaps the most important component that supports successful learning from few examples". In this section we show a real industry scenario (easy generalizable to many other use cases) that we have been working on as part of our company's roadmap. The task has three key ingredients:

1) it's language related;
2) it's in a privacy-aware context (i.e. the system is deployed within an enterprise under security, so no data sharing is possible, even across similar use cases);
3) it's a small-data context: the problem is an NLP (Natural Language Processing) challenge in an enterprise chat, generating a dozen access *per day*.

These ingredients make the problem challenging for a typical DL approach: drawing from ideas in cognitive science (e.g. Goodman et al 2008) and Bayesian inference (e.g. Meylan 2015), we sketch an effective way to frame the problem and start seeing the possibility for a more general solution.

### 3.1 Problem Statement

Consider the following stylized chat-like interface between an A.I. system ("Bot") and one user ("User"):

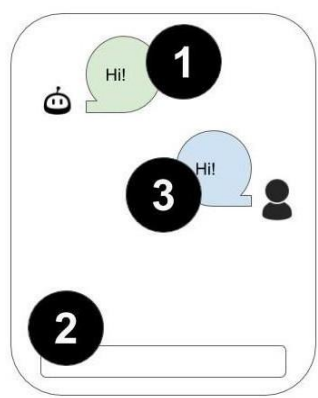

Figure 3: Bot and user interaction.

Bot can send messages to User (as in (1)), and User may reply by typing in the input filed (2); what is typed in (2) can be sent to Bot and be part of the shared conversation between the parties. In the use case at hand, the interface is used by employees inside Company A to ask internal questions about payroll and Human Resources (HR) management - as for example "when are taxes due this year" or "when will I receive my w-2?"[1].

As big as Company A is, this is not Big Data territory, as the amount of data ingested through this interface is fairly limited; moreover, as much as HR lingo is standardized, every organization develops throughout the years "dialects" that are company-specific: this means that the usual "transfer learning" approaches won't be readily applied (Weiss et al 2016).

So, what happens when User chooses a previously unseen word to express a concept (think

---
[1] Form W-2 is an Internal Revenue Service (IRS) tax form used in the United States to report wages paid to employees and the taxes withheld from them (Wikipedia, 2019a).

of Bot being deployed to Company A without previous training on A's lingo)? Compare:

i) 1099 for contractors
ii) 1099 for suppliers
iii) 1099 for externals

where (i) ad (ii) are requests Bot can solve, while (ii) is not since "external" is A's specific synonym for "contractor". How Bot can learn the meaning of "externals" efficiently?[2]

### 3.1.1 Formalizing the Inference with Humans in the Loop

Now that the use case is clear from an industry perspective, let's frame it to cover for the generic case of efficiently learning new lexicon from few examples.

Given a word $w$, a set of observations $X = x_1, x_2 \ldots x_n$, and a set of meaning candidates $E = e_1, e_2 \ldots e_n$ for $w$ (where "meaning" is to be intended as a most general concept, e.g. documents in a search engine, entities in a database, actions in a planning strategy, etc.), a learner (e.g. Bot, in our example) needs to pick the most probable hypothesis $h$ from the set $H$ of functions from $w$ to subsets of $E$. Generically speaking, for any $h_1, h_2 \ldots h_n$ in $H$, Bot can evaluate its posterior probability as:

$$P(h_n|X) \propto P(X|h_n) * P(h_n)$$

that is, the probability of mapping $h_n$ being the meaning for $w$ given data points $X$ is proportional to the *prior* probability of $h_n$ and how well $h_n$ explains the data. Getting back to Bot and User, an interaction may be as follows:

**U**: 1099 for externals

**B**: sorry, I don't know "externals": can you please help me by picking an example from the list below? ["John Contractor", "Company B", "Mike Lawyer"]

**U**: John Contractor

The confidence that "externals" refers to "contractors" for Bot is therefore what can be computed from:

$$P(\text{"external=contractor"}|\textit{John Contractor}) \propto$$
$$P(\textit{John Contractor}|\text{"external=contractor"}) *$$
$$P(\text{"external=contractor"})$$

To sketch a full-fledged solution we then need to specify three things:

1) the prior probability for "external=contractor";
2) the likelihood;
3) how we select the set candidate entities ["John Contractor", "Company B", "Mike Lawyer"] to elicit help from the user.

We will discuss some options for points (1) and (2) in what follows; modelling (3) requires making more precise assumptions on the prior structure, which is out of scope for the current argument.

### 3.1.2 Filling the Slots of the Bayesian Inference

Given a set of entities $E = e_1, e_2 \ldots e_n$, any subset of E is in theory a valid candidate - which means than given $k$ entities in our domain of interest, there are y $= 2^k$ hypotheses, each one with prior probability:

$$P(h) \propto 1 / y$$

Obviously, that is both inefficient and implausible. As knowledge graphs are trending in the industry (Sicular and Brant 2018) and they are independently motivated as chatbot back bone, we shall consider an ontology instead as a way to constrain our space of hypotheses. In particular we shall assume the existence of concepts already partitioning Bot experience of its domain. As an example, consider this small subset of a knowledge graph representing entities and concepts related to the 1099 form:

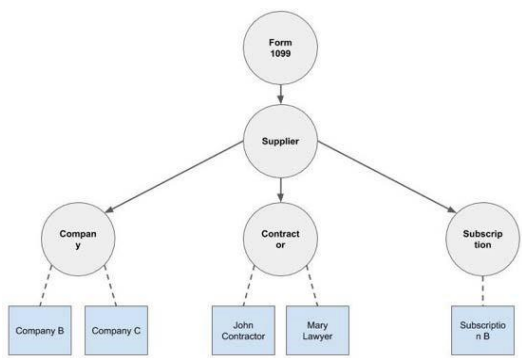

Figure 4: A sample graph showing entities related to the 1099 form[3].

When Bot needs to learn the meaning of "external", the candidates come from the much more constrained structure of the graph. In particular, the prior probability is now spread across three levels of generality: "external" mapping onto the general term "supplier"; "external" mapping onto specific terms, like "company" and "contractor"; finally, no; "external" mapping directly to individuals (in the same sense that, say, "LBJ" and "LeBron James" would map to the same entity). We assign our priors with the intuition that more distinctive concepts are more likely to be distinguished by different words (using the number of siblings plus the node itself as a proxy):

$$P(h) \propto siblings(h) + 1$$

---

[2] Please note that using pre-trained word vectors over big data won't help in this case (and in many similar challenges). For example, Glove vectors provided pre- trained by industry standard libraries, such as (Spacy, 2019), would predict that the closest words to 'external' are 'internal', 'externally', 'input'.

[3] Form 1099 is one of several IRS tax forms used in the United States to prepare and file an information return to report various types of income other than wages, salaries, and tips (Wikipedia, 2019b).

As far as likelihood goes, we can borrow and re-adapt the size principle from (Xu and Tenenbaum, 2007), according to which "smaller hypotheses assign greater likelihood than do larger hypotheses to the same data, and they assign exponentially greater likelihood as the number of consistent examples increases":

$$P(X|h) \propto [1 / ext(h)]^n$$

where ext(h) is the extension of the hypothesis, i.e. the number of objects falling into that category - the number of entities connected to a concept in the graph will be our proxy. Armed with our definitions, we can now plug in the formulas into a probabilistic program and simulate Bot's learning.

## 3.2 An End-to-end Example

As a worked out example of the probabilistic approach we are advocating, these are simulations with the following toy data as related to the challenge of the Company A's chat interface introduced above:

**target word**: "external"
**hypothesis space**:

- external = supplier
- external = company OR external = contractor OR external = subscription
- external = Company B OR external = John Contractor OR ...

**observed data**: ["John Contractor", "Mary Lawyer"]
(e.g. when prompted, User1 selects "John Contractor", User2 selects "Mary Lawyer", etc.)

The image below depicts Bot's probability distribution over the meaning of "external", after the first Bot-User interaction (say, User1), and after the second:

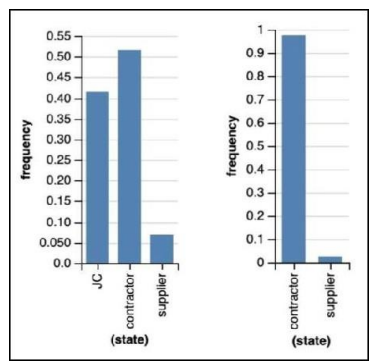

Figure 5: Bot's learning process after one and two user selections: note how quickly the system converges to the correct hypothesis.

We conclude the section with three important observations from the experiment:

1) the proposed probabilistic framework is unique and somewhat in between rule-based and ML approaches: rule-based systems would not learn anything from successive data points from the same concept (as, for example, both "John Contractor" and "Mary Lawyer" are "contractor" and "supplier"); ML approaches instead are great with data points, but they won't be able to make any useful inference after just two observations. The proposed framework gets the best of both worlds leveraging prior knowledge to constrain the search space and data to refine its hypotheses;
2) humans are an essential part of Bot's learning process: when asked by Bot for help, humans cooperation allows the fundamental transfer of conceptual knowledge to the machine; instead of waiting for thousands of interactions, getting humans aligned and involved in the task at end massively speeds up Bot's learning curve. While in use cases such as enterprise chats a great deal of cooperation can be safely assumed, the framework can be extended to handle cases when user selection is noisy (or potentially malicious);
3) all the ingredients respect the hard constraints exposed at the onset of Section 3: Bot learns non- trivial linguistic knowledge; Bot is privacy-aware and does not require any data from Company A to be shared outside the organization; Bot converge very quickly on a reasonable interpretation of user input.

## 4  CONCLUSIONS

It has become commonplace in industry as well as in academia to argue that work is set to disappear through the impact of mass automation and the rise of increasingly more powerful AI (Ford 2015, Poitevin 2017). The picture we have sketched in this article stands in contrast with such a view, though. More precisely, rather than envisage a wholesale replacement of human work, we foresee that a fruitful collaboration between humans and machines can characterize the future of AI.

As argued at length in Section 2, there are good reasons to believe that a great part in the ML and DL industry successes was played by sheer data volume; however, regulatory changes, scientific evidence from human psychology as well business considerations strongly point towards an untapped market for machines that can learn in small data, privacy-aware contexts.

We need to be careful to distinguish between DL and the overall A.I. landscape, which is much more varied than many observers take it to be: as outlined in Section 3 through a fairly general industry use case, there are promising approaches to marry the inference ability of machines with the prior knowledge of humans.

Developing further tools for concept learning is a giant opportunity to deploy scalable A.I. systems *for* humans and *with* humans: if we look at A.I. through the lens of the probabilistic framework we champion, it's easy to see, *pace* Joy 2001, that the future does indeed need us.

## ACKNOWLEDGEMENTS

The authors are immensely grateful for the help of the editors and reviewers in improving and shaping the final version of the article.